\documentclass[12pt]{article}

\usepackage{graphicx}

\begin{document}

\title{\bf Perfect imaging without negative refraction for microwaves}
\author{Yun Gui Ma$^1$, C. K. Ong$^{2}$, Sahar Sahebdivan$^3$, Tom\'{a}\v{s} Tyc$^{3,4}$ and\\ Ulf Leonhardt$^3$\\
$^1$Temasek Laboratories, National University of Singapore,\\ Singapore 119260, Singapore\\
$^2$Centre for Superconducting and Magnetic Materials,\\
Department of Physics, National University of Singapore,\\
Singapore 117542, Singapore,\\
$^3$School of Physics and Astronomy, University of St Andrews,\\
North Haugh, St Andrews KY16 9SS, UK,\\
$^4$Institute of Theoretical Physics and Astrophysics,\\ 
Masaryk University, Kotlarska 2, 61137 Brno, Czech Republic}

\maketitle

\begin{abstract}
We demonstrate perfect imaging in Maxwell's fish eye for microwaves. Our data show that the field of a line source is imaged with subwavelength resolution over superwavelength distances, provided the field is allowed to leave through passive outlets that play the role of a detector array in imaging. 
\end{abstract}

\newpage

Ordinary lenses cannot resolve structures much finer than the wavelength of light \cite{BornWolf}. Perfect lenses made of negatively-refracting metamaterials were predicted \cite{Pendry} to image with unlimited resolution. In practice, however, such materials are absorptive for fundamental reasons \cite{Stockman}; perfect imaging over distances larger than the wavelength seemed impossible. Here we demonstrate imaging with subwavelength resolution over superwavelength distances for microwaves. Like light, microwaves are electromagnetic waves, but with cm wavelengths and GHz frequencies, which allows us to investigate the electromagnetic fields of the imaging waves with a degree of detail currently inconceivable in optics. Instead of using negative refraction, we have implemented a positive refractive-index profile \cite{Maxwell} that appears to curve space for electromagnetic waves \cite{Luneburg,LeoPhil} such that they are focused with infinite precision in principle \cite{Fish}. Our microwave experiment demonstrates the concepts of perfect imaging without negative refraction, in particular the role of detection in achieving perfect resolution, giving important guidance to applications where imaging matters most: for light.

Optical materials may change the spatial geometry perceived by light \cite{LeoPhil}, creating optical illusions such as invisibility \cite{LeoConform,PSS}. Perfect imaging \cite{Pendry,Fish,Fish3D,Spain} is an optical illusion as well where an object appears to be at two or more positions; by perfect imaging we mean the transfer of the electromagnetic field from one place to another, forming a real image at the new location with all the details of the original preserved. For example, negative refraction \cite{Pendry} turns out \cite{GREE} to fold space \cite{CCS}, producing optical ``carbon copies'' on the folded spatial regions. Hyperlenses \cite{Jacob} establish hyperbolic geometries that make magnified virtual images \cite{Liu}. The device we demonstrate here, known as Maxwell's fish eye \cite{Maxwell}, creates the illusion that electromagnetic waves propagate on the surface of a virtual sphere, whereas in reality they are confined to a planar waveguide---any point of the physical plane corresponds to a point on the virtual sphere, a curved space with non-Euclidean geometry \cite{LeoTyc}.

To see why the virtual sphere acts as a perfect imaging device, consider waves continuously emitted from a source in the physical plane and imagine them on the equivalent, virtual sphere. Any source can be regarded as a collection of point sources, so it suffices to investigate the wave produced by a single point source of arbitrary position on the sphere. The wave propagates from the point of emission round the sphere and focuses at the antipodal point (Fig. 1a) that corresponds to the image point in the physical plane. However, the focusing is perfect---infinitely sharp---only if the wave is extracted by an outlet at the image \cite{Fish} (see also the Appendix). By outlet we mean a completely passive, pointlike absorber playing the role of a detector in applications of imaging. Without outlet the wave runs back to the source and forms a stationary pattern lacking subwavelength focus \cite{Comment,Reply}. If only part of the wave manages to escape through the outlet, only that part is perfectly focused. Moreover, we observed in our experiments that when more than one outlet are offered to the wave---as in a detector array---the wave chooses the outlet closest to the correct image point, provided any outlet is within a range from the image point of about half the wavelength. The distance between the detectors may be significantly shorter than the wavelength, so the resolution is subwavelength and, in principle, can become infinitely sharp. Only the detected part of the wave is imaged with point-like precision, but as detection is the very point of imaging, this is perfectly sufficient.

\begin{figure}[h]
\begin{center}
\includegraphics[width=30.0pc]{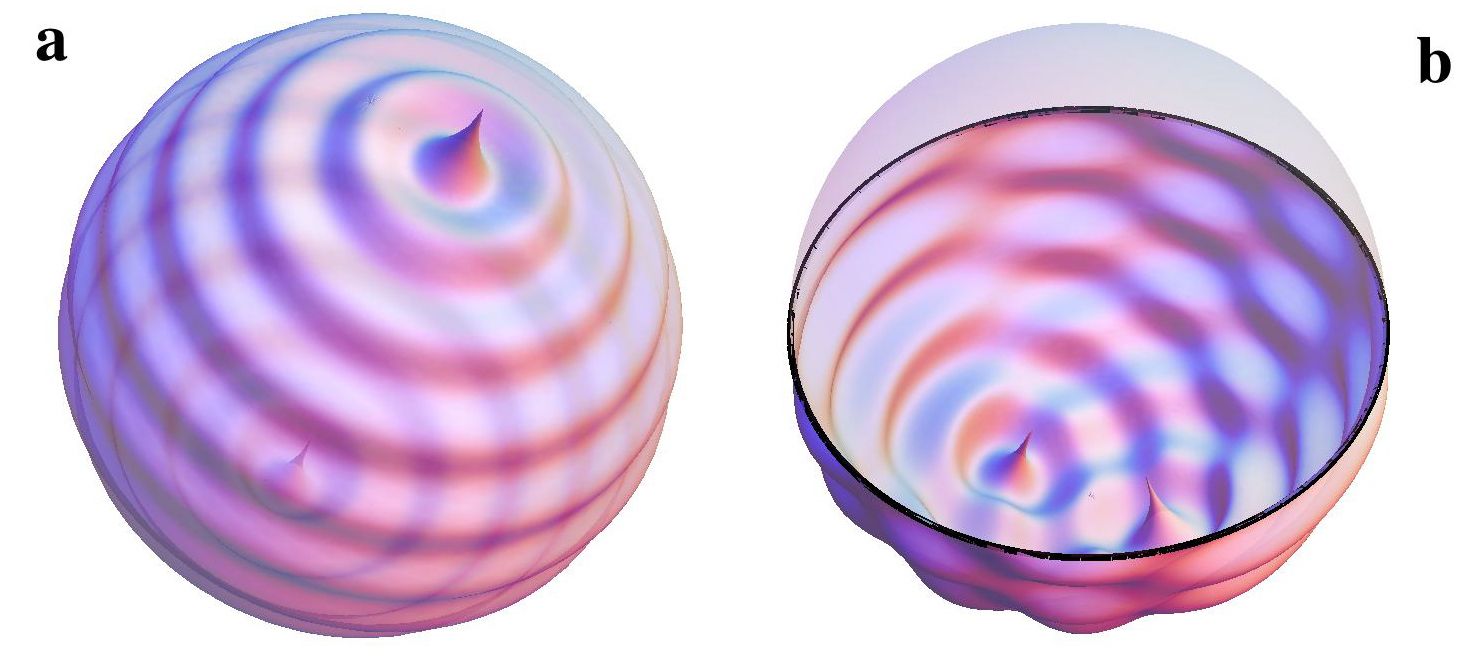}
\caption{
\small{
{\bf Virtual spheres.} {\bf a}: Maxwell's fish eye \cite{Maxwell} creates the illusion that light propagates on the surface of a virtual sphere \cite{Luneburg,Fish}. The wave from a point source (bottom left) propagates round the sphere and focuses at the antipodal point (top right). {\bf b}: a circular mirror is placed around the equator of the virtual sphere such that the wave is focused inside the southern hemisphere.}
}
\end{center}
\end{figure}

Maxwell's fish eye requires a material with refractive index that varies along the distance $r$ from the centre of the device as \cite{Maxwell}
\begin{equation}
n=\frac{2 \, n_1}{1+(r/a)^2} \,.
\end{equation}
Here $a$ is a characteristic length that corresponds to the radius of the virtual sphere; the constant $n_1$ is the refractive index at $r=a$ and also the index on the virtual sphere. In practice, it is advantageous to surround Maxwell's fish eye by a mirror \cite{Fish} at radius $r=a$, which corresponds to a mirror around the equator of the virtual sphere (Fig. 1b). In this case, the index profile ranges from $n_1$ at the mirror to $2 n_1$ in the centre, while still creating perfect images \cite{Fish} (Fig. 1b). Note that Maxwell's fish eye is an unusual ``lens'' where both source and image are inside the device.

Maxwell's fish eye has never been made \cite{Fuchs,Foca}. We have implemented the fish eye mirror \cite{Fish} for microwave radiation confined between two parallel metal plates establishing a planar waveguide \cite{Zhao}. The device is inserted between the plates; its index profile (1) lets microwaves in the planar waveguide behave as if they were waves on a virtual half sphere (Fig. 1b). The plate separation, $5$mm, is chosen such that only microwaves with electric field perpendicular to the plates can travel inside, which is crucial for perfect imaging, because only for electromagnetic waves of this polarization does a material with electric permittivity $\varepsilon=n^2$ appear to curve space perfectly \cite{Fish}. Our device (Fig. 2) resembles a microwave cloaking device \cite{Schurig} or a transmuted Eaton lens \cite{Ma} made of concentric layers of copper circuit board (Rodgers RT6006) with etched-out structures that shape its electromagnetic properties, except that the fish-eye structures respond to the electric and not the magnetic field \cite{Schurig,Ma}. Our structures are designed \cite{Design} for non-resonant operation such that the device can perform perfect imaging over a broad band of the spectrum. For practical reasons, we supplement some layers with dielectric powder (ECCOSTOCK Hik Powders, see Fig. 2); the metal structures and the filling material combined create the desired index profile (1) with $a=5$cm, $n_1=1$. The device has a thickness of $5$mm and fits exactly between the metal plates of the waveguide.

\begin{figure}[t]
\begin{center}
\includegraphics[width=25.0pc]{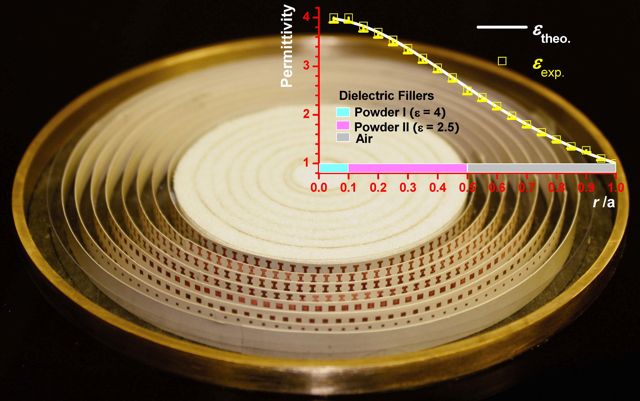}
\caption{
\small{
{\bf The device.} Copper structures on concentric layers of circuit board and dielectric fillers surrounded by a circular metal mirror create the geometry of the half sphere (Fig. 1b) for microwave radiation with electric field pointing in vertical direction. The diagram shows the designed profile \cite{Design} of the electric permittivity $\varepsilon=n^2$ at each layer of the device compared with Maxwell's theoretical formula (1).}
}
\end{center}
\end{figure}

As source we use a coaxial cable inserted through the bottom plate. The cable has an outer diameter of 2.1 mm, $1.68$mm Teflon isolator and $0.5$mm inner conductor; the latter is exposed by $4.5$mm in the device for creating an approximate line source. Through the source cable we inject microwave radiation of free-space wavelength $\lambda_0=3$cm generated by a vector network analyser (HP 8722D) that doubles as synthesiser and analyser. The outlets are inserted through the bottom plate as well, but are completely passive and lead to absorbers impedance-matched to the cables. The outlets are identical to the source such that they act as sources in reverse, for maximal power extraction and best focus \cite{Marques}. The field inside the waveguide is scanned by a coaxial cable inserted through the top plate that is moved in both lateral directions with $1$mm step size \cite{Zhao}. The cable is unexposed such that the field is minimally distorted by the detection. The scanning cable is fed into the vector network analyser where the signal is measured and decomposed into in-phase and out-of-phase components with respect to the synthesised field. Mathematically, these components correspond to the real and imaginary parts of the complex temporal Fourier amplitude taken at each scanned spatial point in the waveguide. 

Figure 3a illustrates the schemes of two experiments for probing the imaging performance of Maxwell's fish eye mirror \cite{Fish}. In the first experiment, we employ one outlet placed at the correct imaging point with respect to the source. In the second experiment, we added another outlet placed at $0.2\lambda$ distance from the first outlet where $\lambda=\lambda_0/n$ is the local wavelength at the image. Figure 3b displays the scanned field intensity (the modulus squared of the complex Fourier amplitude) clearly showing subwavelength focusing at the image spot. When the second outlet is added the intensity profile is nearly identical (Fig. 3c), which proves that the wave is focused at the correct outlet, even when the outlets are closer than the standard diffraction limit \cite{BornWolf} of $0.5\lambda$.  

\begin{figure}[h]
\begin{center}
\includegraphics[width=35.0pc]{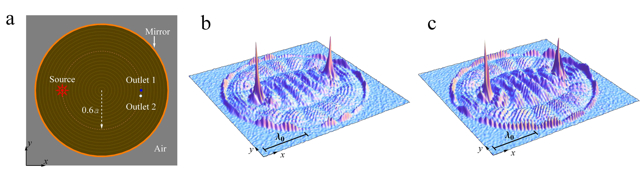}
\caption{
\small{
{\bf Experiments.} {\bf a}: Scheme of two experiments where microwaves may run from the source to one or two passive outlets that play the role of detectors. {\bf b}: modulus squared of the scanned electric-field amplitude for the case when the outlet is placed at the correct image point; $\lambda_0$ indicates the free-space wavelength. {\bf c}: a second outlet is added at subwavelength distance from the first one. The two intensity profiles show sharp peaks at the correct image point that are nearly indistinguishable, proving that the radiation goes into the right outlet/detector with subwavelength resolution. }
}
\end{center}
\end{figure}

Figure 4 compares the field in the first experiment, scanned along the line between source and image, with the theoretical prediction \cite{Fish} based on assuming the perfectly smooth index profile (1) and ideal line sources. The figure shows both the real and the imaginary part of the field amplitude, thus proving that most of the injected microwave radiation establishes a running wave \cite{Reply} that leaves the device at the outlet. The agreement with theory \cite{Fish} is remarkably good, considering that the device is made of a structured material and that source and outlet are not ideal. The source launches electromagnetic waves that, in the near field, have also electric components parallel to the plates, and source and outlet have electromagnetic cross sections much larger than their geometrical size \cite{Combleet,Jackson}. It seems that at present the imaging resolution is limited by the source and the detector, which, in principle can be made perfect.

{\bf Acknowledgements.}
Y.G.M. and C.K.O. are supported by DSO National Laboratories and the Defence Research and Technology Office (DRTech) of Singapore, T.T. acknowledges the grants MSM 0021622409 and MSM 0021622419 and U.L. is supported by a Royal Society Wolfson Research Merit Award and a Theo Murphy Blue Skies Award of the Royal Society.

\begin{figure}
\begin{center}
\includegraphics[width=25.0pc]{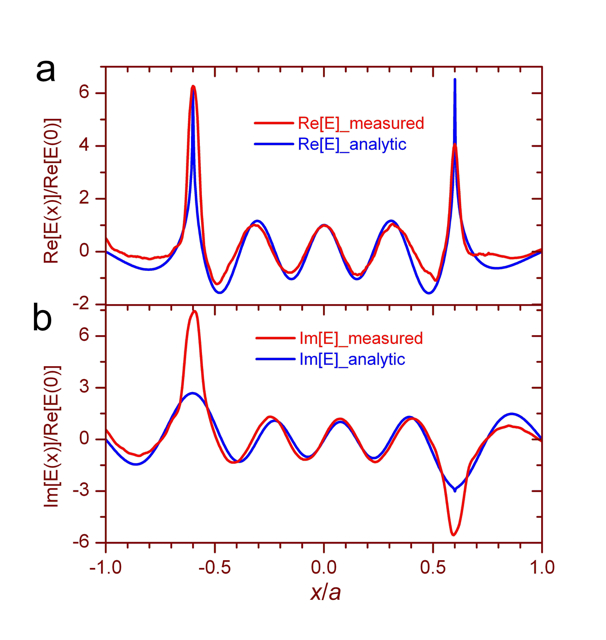}
\caption{
\small{
{\bf Comparison with theory.} The field amplitude scanned along the line between source and outlet is compared with the analytical formula of a theory \cite{Fish} where the perfect index profile (1) and an infinitely localized line source and outlet were assumed. {\bf a}, real part, {\bf b}, imaginary part of the complex Fourier amplitude. The figure shows a running wave \cite{Reply} in good agreement with theory \cite{Fish}; the deviations from theory are primarily due to imperfections in source and outlet.}
}
\end{center}
\end{figure}

\newpage

\noindent
{\bf \Large Appendix}\\

\renewcommand{\theequation}{A\arabic{equation}}
\setcounter{equation}{0}

In this appendix we compile the analytic expressions we used for comparing our microwave data with theory and we show experimental results for imaging without outlet and hence without subwavelength resolution.

For simplicity, we describe the Cartesian coordinates $x$ and $y$ in the plane of the waveguide in units of the device radius $a$ and we put $n_1=1$. It is convenient to combine the two coordinates in one complex number
$z=x+\mathrm{i}z$. In this notation and with our units, the refractive-index profile of Maxwell's fish eye \cite{Maxwell} reads
\begin{equation}
n = \frac{2}{1+|z|^2} \,.
\label{max}
\end{equation}
Consider stationary electromagnetic waves with wavenumber $k$ (in our units) and electric field polarized in vertical direction. In this case the electric-field strength is characterized by only one scalar complex Fourier amplitude $E$ that depends on $k$ and $z$; we denote it by $E_k(z)$. We assume that the wave propagates inside a material with electric permittivity $\varepsilon=n^2$ and index profile (\ref{max}) surrounded by a perfect mirror at $r=1$. Theory \cite{Fish} shows that the field of a perfect line source is given by the exact expressions 
\begin{equation} 
E_k(z) =  E_{\nu} (z) - E_{\nu} (1/z^*) \,,\quad E_{\nu} = \frac{P_{\nu} (\zeta) -  \mathrm{e}^{\mathrm{i} \nu \pi} \, P_{\nu} (-\zeta) }{4 \sin(\nu\pi)} \,,
\label{field}
\end{equation}
where the $P_\nu$ are Legendre functions \cite{Erdelyi} with the index
\begin{equation}
\nu = \frac{1}{2}\left(\pm\sqrt{4 k^2 + 1} - 1\right) .
\label{nu} 
\end{equation}
The plus sign refers to positive wavenumbers $k$ and the minus sign to negative $k$ (we shall need negative $k$ for describing the field in the case without outlet). For the variable $\zeta$ of the Legendre functions we have
\begin{equation}
\zeta = \frac{|z'|^2 - 1}{|z'|^2 + 1}  \,,\quad
z'= \frac{z-z_0}{z_0^*z + 1} \,,
\end{equation}
where $z_0$ denotes the coordinates $x_0$ and $y_0$ of the line source in complex representation $z_0=x_0+\mathrm{i}y_0$. The wave function (\ref{field}) develops two logarithmic singularities \cite{Fish} within the region $|z|<1$ of the device, one at the source $z_0$ and one at the image point
\begin{equation}
z_0' = -z_0 \,.
\label{image}
\end{equation}
This means that the wave forms an exact image with, in principle, unlimited resolution. The singularity at the image turns out \cite{Fish} to carry the phase factor $\exp(\mathrm{i}\pi\nu)$, so the phase delay is $\pi\nu$. Figure 4 of our paper shows that expression (\ref{field}) agrees well with our data, apart from imperfections due to the finite electromagnetic size of source and image. Note that formula (\ref{field}) describes the field of a running wave that disappears through the outlet at the focal point and forms a perfect image. This outlet is a completely passive absorber that plays the role of a detector in imaging. 

In the case when no outlet is present, the wave runs back to the source where it is reabsorbed, establishing a stationary wave. Imagine the stationary wave as a continuous stream of elementary flashes of radiation. Near the image (\ref{image}) each elementary wave focuses like the radiation emitted by the source run in reverse, like an advanced solution \cite{Jackson} of Maxwell's equations, but when the wave runs back it appears like a retarded wave \cite{Jackson}. Therefore, the total electromagnetic wave in the stationary regime without outlet is the superposition of an advanced and retarded wave \cite{Reply}
\begin{equation}\label{super}
E'_k(z) = \frac{E_k(z)-\mathrm{e}^{\mathrm{i}\pi\nu(k)-\mathrm{i}\pi\nu(-k)}E_{-k}(z)}{1-\mathrm{e}^{\mathrm{i}\pi\nu(k)-\mathrm{i}\pi\nu(-k)}} \,.
\end{equation}
One verifies that expression (\ref{super}) describes a real field with logarithmic singularity at the source, as required \cite{Reply}. The real field (\ref{super}) forms a standing wave like the plane wave $\cos(kx)$ in free space, in contrast to the wave (\ref{field}) that is complex and corresponds to a running wave like $\exp(\mathrm{i}kx)$. One also verifies that the standing wave (\ref{super}) does not develop a singularity at the image point (\ref{image}): the standing wave (\ref{super}) does not form a perfect image.

\begin{figure}
\begin{center}
\includegraphics[width=32.0pc]{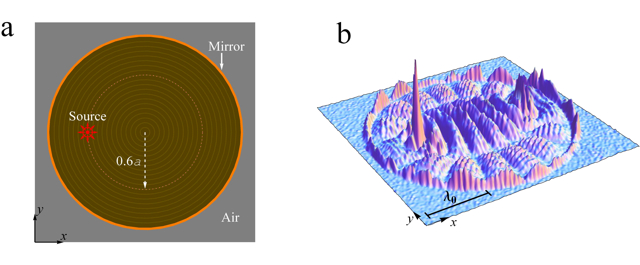}
\caption{
\small{
{\bf Imaging without outlet.} {\bf a}: Scheme of the experiment: microwaves are emitted and reabsorbed by the source; no outlet is present. {\bf b}: modulus squared of the scanned electric-field amplitude; no sharp image is formed.}
\label{fig:field}
}
\end{center}
\end{figure}

\begin{figure}
\begin{center}
\includegraphics[width=25.0pc]{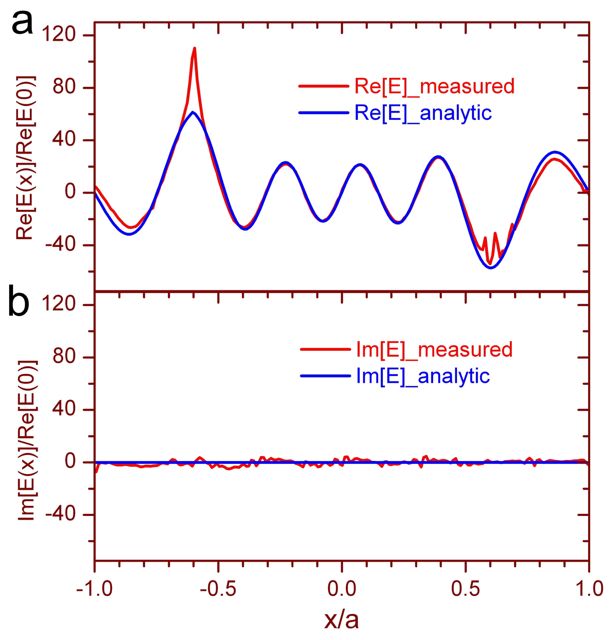}
\caption{
\small{
{\bf Comparison with theory.} The field amplitude scanned along the line between source and outlet is compared with the analytical formula (\ref{super}). The figure shows a standing wave \cite{Reply} in very good agreement with theory \cite{Fish}; the subwavelength features near the image originate from the structures of the material used to implement the fish eye mirror, the deviation near the source is due to its imperfection.}
\label{fig:compare}
}
\end{center}
\end{figure}

Figure~\ref{fig:field} shows our experimental results for imaging without outlet, when no detector monitors the field. Instead of the sharp peak in perfect imaging with outlet (Fig.~3b of our paper) the wave forms a diffraction-limited focus. Figure~\ref{fig:compare} compares the measured field with formula (\ref{super}). One sees that without outlet the wave is real and so a standing wave is formed. Here theory and experiment agree even better than in the perfectly-imaging regime, because the experimental situation is simpler; the wave is not required to escape through the outlet. The subwavelength features near the image originate from the structure of the material used to implement Maxwell's fish eye, the rings of circuit board (Fig.~2). As each elementary wave of the continuous radiation attempts to focus there with perfect precision before being reflected back to the source, the subwavelength structure of the device near the image becomes apparent. Our experimental results show that only the detected field is perfectly imaged in Maxwell's fish eye \cite{Maxwell}.

\end{document}